\documentclass[sigconf]{acmart}

\AtBeginDocument{%
  }

\copyrightyear{2024}
\acmYear{2024}
\setcopyright{rightsretained}
\acmConference[AM '24]{Audio Mostly 2024 - Explorations in Sonic Cultures}{September 18--20, 2024}{Milan, Italy}
\acmBooktitle{Audio Mostly 2024 - Explorations in Sonic Cultures (AM '24), September 18--20, 2024, Milan, Italy}
\acmPrice{}
\acmDOI{10.1145/3678299.3678328}
\acmISBN{979-8-4007-0968-5/24/09}

\usepackage[utf8]{inputenc} 
\usepackage[T1]{fontenc} 
\usepackage{amsmath,url}
\usepackage{graphicx} 
\usepackage{multirow}
\usepackage{diagbox}
\usepackage{color}
\usepackage{balance}
\usepackage{colortbl}
\usepackage{hhline}
\usepackage{url}

\usepackage{hyperref}

\title{Joint Learning of Emotions in Music and Generalized Sounds}
\author{Federico Simonetta}
\orcid{0000-0002-5928-9836}
\affiliation{%
  \institution{GSSI - Gran Sasso Science Institute}
  \city{L'Aquila}
  \country{Italy}
}
\email{federico.simonetta@gssi.it}

\author{Francesca Certo}
\orcid{0000-0002-5082-5903}
\email{francesca.certo@studenti.unimi.it}
\author{Stavros Ntalampiras}
\orcid{0000-0003-3482-9215}
\email{stavros.ntalampiras@unimi.it}
\affiliation{%
  \institution{University of Milan}
  \city{Milan}
  \country{Italy}}

\renewcommand{\shortauthors}{Simonetta et al.}

\begin{document}


\begin{abstract}
  In this study, we aim to determine if generalized sounds and music can share a common emotional space, improving predictions of emotion in terms of arousal and valence.
  We propose the use of multiple datasets as a multi-domain learning technique. Our approach involves creating a common space encompassing features that characterize both generalized sounds and music, as they can evoke emotions in a similar manner. To achieve this, we utilized two publicly available datasets, namely IADS-E and PMEmo, following a standardized experimental protocol. We employed a wide variety of features that capture diverse aspects of the audio structure including key parameters of spectrum, energy, and voicing. Subsequently, we performed joint learning on the common feature space, leveraging heterogeneous model architectures. Interestingly, this synergistic scheme outperforms the state-of-the-art in both sound and music emotion prediction. The code enabling full replication of the presented experimental pipeline is available at \url{https://github.com/LIMUNIMI/MusicSoundEmotions}.

\end{abstract}


\maketitle

\section{Introduction}
Emotions have always played a fundamental role in human lives, and they are currently receiving increasing attention in the technological field~\cite{8816973}. Although emotions and computer science have traditionally been viewed as two distinct concepts due to the lack of consciousness in computers, which prevents them from experiencing emotions, numerous studies have been conducted over the years to demonstrate computers’ ability to identify people’s moods and emotions.



The present research aims to determine whether generalized sounds and music can share a common emotional space. The study proposes a novel multi-domain learning approach that harnesses the power of affective computing. Utilizing datasets representing both music and general sounds, which are capable of eliciting comparable emotional responses, the study creates a shared feature space for emotion prediction. Traditionally distinct audio domains converge in this research, suggesting that a unified model can offer enhanced performance in interpreting emotional responses to a broad spectrum of auditory stimuli.


Affective Computing utilizes two
primary emotion representation frameworks: categorical (e.g., happiness, anger) and
dimensional (e.g., arousal, valence). In this study, we will specifically focus on datasets that provide annotations of perceived emotions in the continuous space of valence and arousal.



The study of computational techniques for analyzing and recognizing emotions in sounds
is known as Audio Emotion Recognition (AER), which is a subfield of Affective Computing~\cite{8816973}. While speech and music have been extensively studied in the literature, recent research has also explored general sound events that may impact human emotional states. 

Music can convey emotions through melody and lyrics, thus affecting the emotional state of listeners. Music Emotion Recognition (MER) systems have been developed for various applications, including medical applications to improve patients' physical and mental health
and music players to recommend songs based on the user's mood. 
Zhang et al.~\cite{Zhang:2018:PDM:3206025.3206037} introduced the PMEmo dataset, extensively studied for music emotion recognition (MER) using multimodal or audio-only approaches.
Among the various works about MER, three publications used PMEmo with a strong and accurate validation pipeline. First,  De Berardinis et al.~\cite{deberardinis2020MultipleVoicesMusical} introduced EmoMucs, a computational model that considers the role of different musical voices in predicting emotions induced by music. Second, Chowdhury et al.~\cite{chowdhury2021TracingBackMusic} proposed a method to trace music emotion predictions back to sound sources and intuitive perceptual qualities. Third, Huang et al.~\cite{huang2022ADFFAttentionBased} proposed an end-to-end attention-based deep feature fusion approach for MER. 


Despite the importance of sound events in individuals' daily lives, research on emotion prediction of sound events has received less attention when compared to speech and music. 
Previous works investigated the analogies between music, speech, and sounds in the context of emotion recognition. Weninger et al.~\cite{weninger2013acoustics}, Ntalampiras~\cite{ntalampiras2017transfer}, and Coutinho et al.~\cite{coutinho2014transfer} all studied various techniques, such as analyzing analogies between speech, music, and sound events, constructing shared emotional spaces, and employing transfer learning to improve the prediction of emotion in music and speech. 
In a seminal work, Bradley and Lang~\cite{bradley2007iads} developed the widely used International Affective Digitized Sounds (IADS) dataset, which has been extended~\cite{yang2018AffectiveAuditoryStimulus} and utilized in various studies. 
Abri et al.~\cite{abri2021ComparativeAnalysisModeling} developed machine and deep learning models to predict the emotions associated with certain sounds, and compared the accuracy of those predictions using IADS-E based on a well-defined validation strategy.

This study focuses on acoustic stimuli in the form of environmental sounds, noises, and music. We investigate the ability of regression models to predict dimensional emotions while utilizing different types of sounds. Specifically, we analyze two types of sounds: \textit{music} and \textit{environmental sounds}. Our experimental results demonstrate that training models with both music and generic sounds lead to more robust models and more accurate emotion predictions for both types of sounds. The expectation is that, by capturing intricate emotional elements across diverse sound types, the approach outlined in this research stands to contribute appreciably to the practical and theoretical advancements in affective computing as it pertains to the world of the Internet of Sounds.

The primary contributions of this study are as follows: (i) a novel multi-modal learning strategy for Audio Emotion Recognition (AER) models combining two different types of sounds, (ii) new models that surpass the state-of-the-art in emotion recognition for both music and environmental sounds, and (iii) an accurate analysis of the impact of the proposed augmentation strategy on the two types of sounds.

The remainder of this paper is organised as follows: section \ref{sec:methodology} explains the proposed methodology including the datasets, feature extraction and modeling process. Section \ref{sec:experiments} presents the experimental set-up and analysed the obtained resutls, while section \ref{sec:conclusions} summarizes the main findings. 

\begin{figure}[t]
    \includegraphics[width=0.5\textwidth]{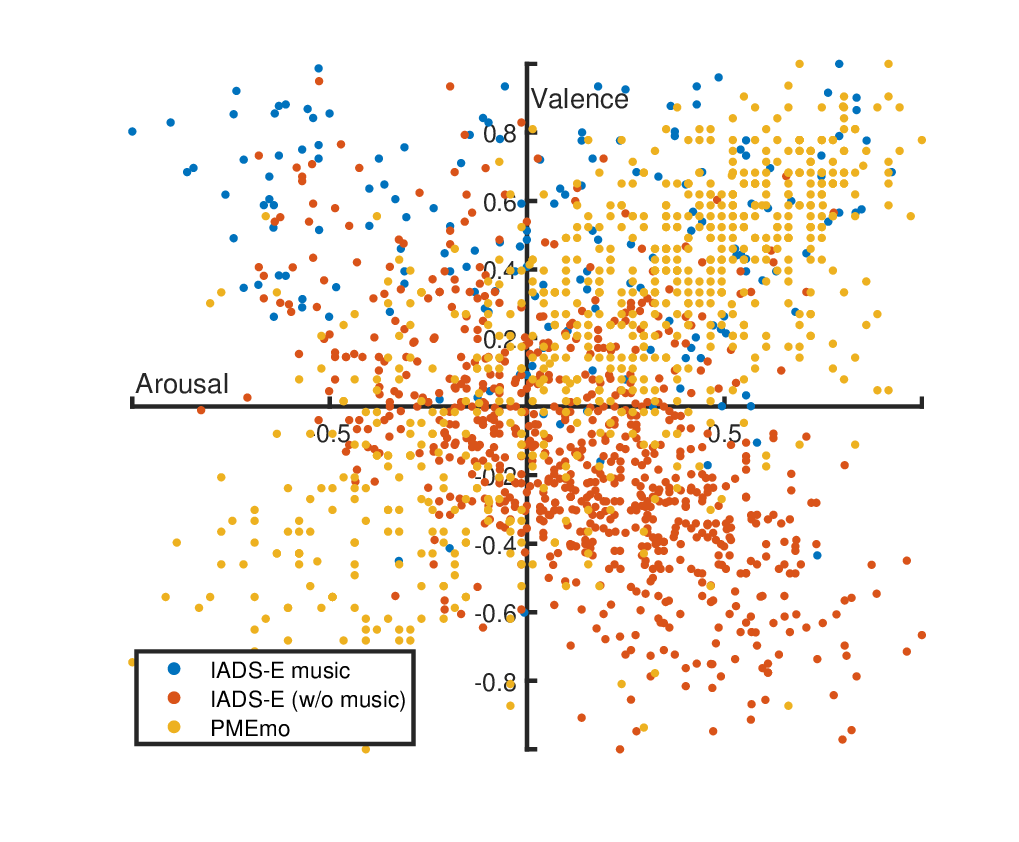}  
    \Description{Image}\caption{Distribution of the ratings in both datasets on the valence-arousal plane.}
  \label{fig:distribution}
\end{figure}

\section{Methodology}
\label{sec:methodology}
The approach in this study analyzes the efficacy of various regression models in predicting emotions evoked by general sound events and music. The models were trained using both single and combined data sets.

Specifically, this study aims to showcase the efficacy of combining two distinct types of sounds in enhancing emotion prediction accuracy. This finding underscores the presence of certain audio characteristics that elicit comparable emotional responses in individuals, regardless of the domain of sound. Moreover, the proposed approach capitalizes on the availability of diverse datasets to offer a straightforward, yet potent augmentation strategy.

\begin{figure*}
    \centering
    \includegraphics[width=\textwidth]{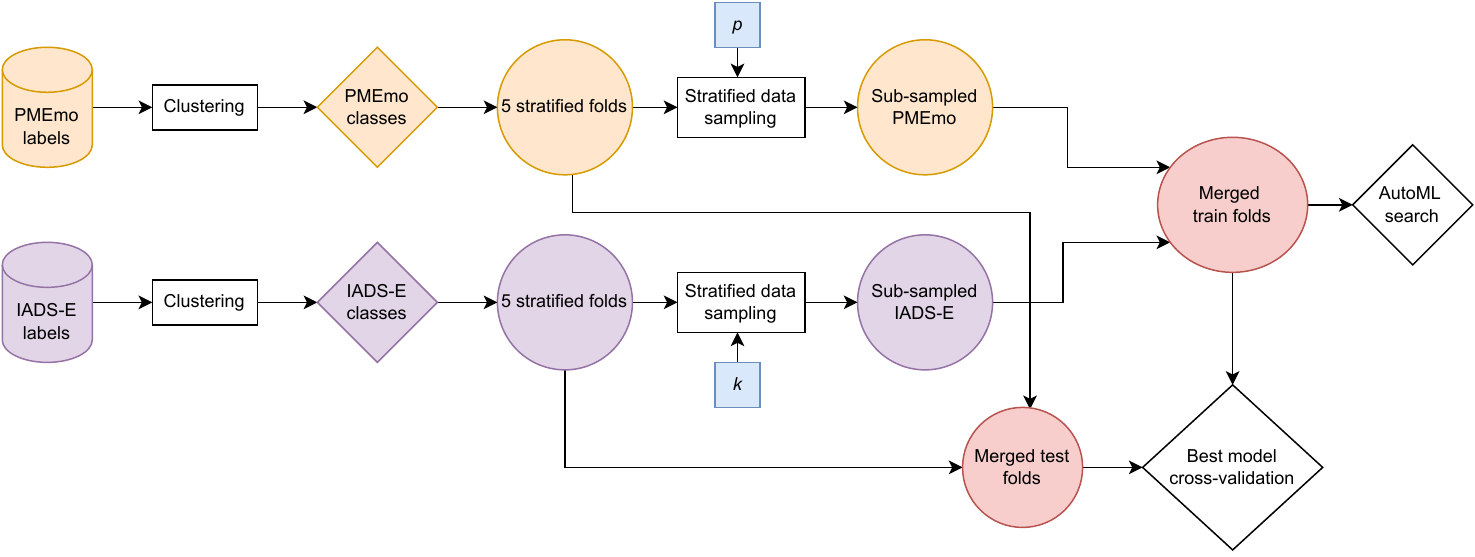}
    \Description{Image}\caption{The overall pipeline: first, clustering is suitably used for applying stratified sampling; then, datasets are sub-sampled according to the parameters $k$ and $p$; finally, the model learns the merged sub-populations and is tested on the original test folds.}
    \label{fig:diagram}
\end{figure*}

The overall pipeline involves clustering the valence-arousal labels and sub-sampling the datasets based on the distribution of the samples across the clusters. Then, 5-fold cross-validation is accurately used to compare the impact of merging different data domains on the accuracy of AutoML pipelines. The general pipeline is represented in Fig.~\ref{fig:diagram}.

\subsection{Data sets}


This work investigates the behavior of models trained on specific sound types when applied to different ones. We aim to determine if two classes of sounds can share a common emotional space, improving predictions of emotion in terms of arousal and valence. We used the IADS-E and PMEmo datasets to analyze the emotional space of sound events and music, respectively.


IADS-E~\cite{yang2018AffectiveAuditoryStimulus} expands the existing auditory affective sample database. It includes ratings from 207 participants using the SAM and basic-emotion rating scales on 935 sounds, including those from IADS-2~\cite{bradley2007iads}. The results showed that emotions in sounds can be distinguished on affective rating scales, providing a larger corpus of natural, emotionally evocative auditory stimuli covering a wide range of categories. IADS-E also provides a semantic categorization based on labels assigned by 10 additional users, comprising 10 classes.
While the IADS-E dataset constitutes a large set of sounds comprehensive of several categories that are rare in other existing datasets, it is limited by the asbence of speech sounds and utterances. Nevertheless, in our study, these categories are partially covered by music recordings.


The PMEmo dataset~\cite{Zhang:2018:PDM:3206025.3206037} consists of 794 popular music choruses, with 767 annotated with static emotion labels in terms of arousal and valence. It is designed for research on Music Emotion Recognition and Music Information Retrieval. Similar to the IADS-E dataset, the SAM technique was used to annotate emotional experiences along arousal and valence dimensions. In this study, we used static arousal and valence ratings to train regression models.


To standardize ratings and aid regression model learning, labels in IADS-E and PMEmo datasets were rescaled to $[-1, 1]$. Fig. \ref{fig:distribution} shows ratings of IADS-E music, IADS-E without music and PMEmo, illustrating their complementarity. No single dataset covers all quadrants of the valence-arousal plane, suggesting a holistic approach when learning from a combined dataset.

\subsection{Feature Extraction}

Feature extraction plays a crucial role in Affective Computing as it transforms audio samples into numerical features that can be processed by machine learning algorithms without compromising the original information. This stage is paramount in determining the relevant features to be used as input in predictive models, ultimately influencing the accuracy of emotion prediction.

In this study, we utilized the \texttt{openSMILE} toolkit~\cite{schuller2014INTERSPEECH2014Computational} to extract audio features from the IADS-E and PMEmo datasets. The \texttt{openSMILE} toolkit integrates features from both speech and music domains, providing a versatile software for feature extraction that is domain-independent. 

For this purpose, we employed the ComParE 2013 configuration~\cite{schuller2014INTERSPEECH2014Computational} packaged with version 3.0.1 of the \texttt{openSMILE} toolkit. The extracted feature set captures essential parameters of spectrum, energy, and voicing. Furthermore, ComPaRe applies various statistical functions to these low-level descriptors to capture diverse aspects of their temporal evolution. In total, 6375 static features are extracted for each sound sample using ComPaRe. 

\begin{table*}[t]
  \centering
  \caption{Prediction results in terms of RMSE achieved by the considered model types when using a) PMEmo, b) IADS-E, and c) the fully augmented dataset.}
  \label{tab:rmse}
  \begin{tabular}{|lll|l|l|l|}
  \hline
  \multicolumn{3}{|l|}{\diagbox{Test set}{Train set}}                                                                                                            & \textit{\textbf{IADS-E (no music)}} & \textit{\textbf{PMEmo}} & \textit{\textbf{IADS-E (no music) + PMEmo}} \\ \hline
  \multicolumn{1}{|l|}{\multirow{6}{*}{\textit{\textbf{IADS-E (no music)}}}} & \multicolumn{1}{l|}{\multirow{2}{*}{\textit{\textbf{Linear}}}} & \textit{Arousal} & \textbf{2.14e-01 ± 2.06e-02}        & 2.80e+04 ± 3.54e+04     & 1.69e+00 ± 4.08e+00                         \\ \cline{3-6} 
  \multicolumn{1}{|l|}{}                                                     & \multicolumn{1}{l|}{}                                          & \textit{Valence} & 3.06e-01 ± 1.98e-02                 & 3.60e+04 ± 2.80e+04     & \textbf{2.61e-01 ± 1.25e-02}                \\ \cline{2-6} 
  \multicolumn{1}{|l|}{}                                                     & \multicolumn{1}{l|}{\multirow{2}{*}{\textit{\textbf{SVM}}}}    & \textit{Arousal} & 1.92e-01 ± 1.03e-02                 & 2.66e-01 ± 1.96e-03     & \textbf{1.25e-01 ± 1.15e-02}                \\ \cline{3-6} 
  \multicolumn{1}{|l|}{}                                                     & \multicolumn{1}{l|}{}                                          & \textit{Valence} & 3.58e-01 ± 9.42e-03                 & 3.62e-01 ± 4.82e-03     & \textbf{2.41e-01 ± 2.87e-02}                \\ \cline{2-6} 
  \multicolumn{1}{|l|}{}                                                     & \multicolumn{1}{l|}{\multirow{2}{*}{\textit{\textbf{AutoML}}}} & \textit{Arousal} & 1.85e-01 ± 1.38e-02                 & 2.45e-01 ± 1.17e-02     & \textbf{1.04e-01 ± 5.93e-03}                \\ \cline{3-6} 
  \multicolumn{1}{|l|}{}                                                     & \multicolumn{1}{l|}{}                                          & \textit{Valence} & 2.71e-01 ± 1.15e-02                 & 3.24e-01 ± 2.25e-02     & \textbf{2.53e-01 ± 1.31e-02}                \\ \hline
  \multicolumn{1}{|l|}{\multirow{6}{*}{\textit{\textbf{PMEmo}}}}             & \multicolumn{1}{l|}{\multirow{2}{*}{\textit{\textbf{Linear}}}} & \textit{Arousal} & 6.17e-01 ± 1.87e-01                 & 2.92e-01 ± 8.19e-02     & \textbf{2.29e-01 ± 2.37e-02}                \\ \cline{3-6} 
  \multicolumn{1}{|l|}{}                                                     & \multicolumn{1}{l|}{}                                          & \textit{Valence} & 1.32e+00 ± 4.11e-01                 & 4.45e-01 ± 4.72e-01     & \textbf{2.41e-01 ± 2.87e-02}                \\ \cline{2-6} 
  \multicolumn{1}{|l|}{}                                                     & \multicolumn{1}{l|}{\multirow{2}{*}{\textit{\textbf{SVM}}}}    & \textit{Arousal} & 3.62e-01 ± 1.46e-02                 & 2.10e-01 ± 4.88e-03     & \textbf{1.93e-01 ± 9.23e-03}                \\ \cline{3-6} 
  \multicolumn{1}{|l|}{}                                                     & \multicolumn{1}{l|}{}                                          & \textit{Valence} & 4.14e-01 ± 1.41e-02                 & 2.55e-01 ± 2.09e-02     & \textbf{1.68e-01 ± 2.10e-02}                \\ \cline{2-6} 
  \multicolumn{1}{|l|}{}                                                     & \multicolumn{1}{l|}{\multirow{2}{*}{\textit{\textbf{AutoML}}}} & \textit{Arousal} & 3.20e-01 ± 9.28e-03                 & 1.93e-01 ± 7.22e-03     & \textbf{1.80e-01 ± 1.53e-02}                \\ \cline{3-6} 
  \multicolumn{1}{|l|}{}                                                     & \multicolumn{1}{l|}{}                                          & \textit{Valence} & 3.89e-01 ± 3.16e-02                 & 2.23e-01 ± 2.12e-02     & \textbf{1.53e-01 ± 2.62e-02}                \\ \hline
  \end{tabular}%
\end{table*}

\subsection{Model Selection and Validation on Combined Data Sets}

To evaluate the impact of the proposed augmentation strategy on different models, we employed three approaches. Firstly, we utilized a linear model based on ElasticNet and a Support Vector Regression (SVR) model as implemented in the \texttt{sklearn} library. Secondly, we adopted a state-of-the-art AutoML method to evaluate the impact of various pre-processing and feature selection strategies on model performance. The AutoML also enabled us to compare a variety of regressors, both linear and non-linear, and to create an ensemble of the best performing models~\cite{feurer2022AutosklearnHandsfreeAutoML}.

\begin{table*}
    \centering
      \caption{Ablation study demonstrating how the proposed learning scheme compares with the state of the art.}
      \label{tab:literature}
        \begin{tabular}{|l|l|l|l|l|l|}
            \hline
            \multicolumn{1}{|l|}{\multirow{2}{*}{\textbf{Test set}}} & \multirow{2}{*}{\textbf{Method}}                       & \multicolumn{2}{l|}{\textit{\textbf{RMSE}}}                                   & \multicolumn{2}{l|}{\textbf{$R^2$}}                       \\ \cline{3-6} 
            \multicolumn{1}{|l|}{}                                   &                                                        & \multicolumn{1}{l|}{\textit{Valence}} & \multicolumn{1}{l|}{\textit{Arousal}} & \multicolumn{1}{l|}{\textit{Valence}} & \textit{Arousal}  \\ \hline
            \multirow{3}{*}{\textit{\textbf{IADS-E (with music)}}}   & \textit{Abri et al., 2021}                             & .289 *                            & .195 *                            & .370                              & .563          \\
                                                                     & \textit{AutoML on IADS-E (with music)}         & .279                              & .193                              & .379                              & .566          \\
                                                                     & \textit{AutoML  on IADS-E (with music) + PMEmo} & \textbf{.249}                     & \textbf{.163}                     & \textbf{.508}                     & \textbf{.692} \\ \hline
            \multirow{6}{*}{\textit{\textbf{PMEmo}}}                 & \textit{de Berardinis et al., 2020}                    & .232                              & .223                              & .481                              & .610          \\
                                                                     & \textit{Chowdhury et al., 2021}                        & .310                              & .250                              & .400                              & .600          \\
                                                                     & \textit{Huang et al., 2022}                            & .231                              & .216                              & .508                              & .655          \\
                                                                     & \textit{AutoML  on PMEmo}                       & .223                              & .193                              & .525                              & .727          \\
                                                                     & \textit{AutoML  on IADS (no music) + PMEmo}     & .153                              & .180                              & .775                              & .762          \\
                                                                     & \textit{AutoML  on IADS-E (with music) + PMEmo} & \textbf{.152}                     & \textbf{.137}                     & \textbf{.780}                     & \textbf{.861} \\ \hline
            \end{tabular}
            
            * These values were re-normalized so that they are referred to labels in $[-1, 1]$
            
            \textit{Note}: all the results are referred to 5-fold cross-validation and labels normalized in $[-1, 1]$; except for the proposed model, for papers showing multiple configurations of the proposed models, the best results were picked, even if produced by different model configurations.
\end{table*}

Both ElasticNet and SVR were trained using the principal components (PCs) of the features extracted from the IADS-E and PMEmo datasets. The number of PCs was determined by the cumulative explained variance ratio, which was itself optimized as a hyper-parameter of the model.
The optimization of ElasticNet and SVR was performed using a successive halving grid search. Both the AutoML and the successive halving grid search were executed on a single machine with 32 GB of RAM and 12 cores.
For the sake of space, the code documentation provides detailed information on the hyperparameter space for ElasticNet and SVR. Both AutoML and grid-search optimized the mean squared error (MSE).

To construct a mixed dataset, we merged the PMEmo and IADS-E datasets in proportions governed by two parameters, which dictate the respective dataset contributions to the mix. we define the combined dataset size using the equation: 

\begin{equation}
    k \times |\textit{IADS-E}| + p \times |\textit{PMEmo}|,
\end{equation}

where $k, p \in [0, 1]$. Here, $|\textit{IADS-E}|$ and $|\textit{PMEmo}|$ represent the total number of samples in the PMEmo and IADS-E datasets, respectively. We constrain the parameters such that either $k$ or $p$ is set to $1$, ensuring that one dataset is fully included while the inclusion of the other is adjustable.
For a given pair $(k, p)$, both AutoML and successive halving were utilized with 5-fold cross-validation. The hyper-parameter optimization was conducted to search for models that performed well \textit{on average} for both PMEmo and IADS-E when trained on the augmented dataset.

We then selected the best-performing model from the hyper-parameter optimization and the best-performing ensemble from AutoML, and evaluated them using a similar 5-fold cross-validation procedure. Interestingly, this time we observed the performance on the validation fold of IADS-E and PMEmo separately.

The cross-validations were performed using a stratified sampling in order to address the potential issue of overfitting due to non-equitable representation of the dataset across the folds. Since, the target labels of the datasets at hand are continuous, we employed the Ward hierarchical method for clustering and then associated a class to each cluster. The same approach was used for sub-sampling the datasets, i.e. when $k$ and $p$ are not set to $1$.
To ensure an adequate number of samples in each fold, we set the minimum number of samples in a cluster to 25 as our stopping criterion.

\section{Experimental Set-Up and Results}
\label{sec:experiments}

\begin{figure*}[t]
  \begin{center}
    \includegraphics[width=0.45\textwidth, trim=0 40 0 0]{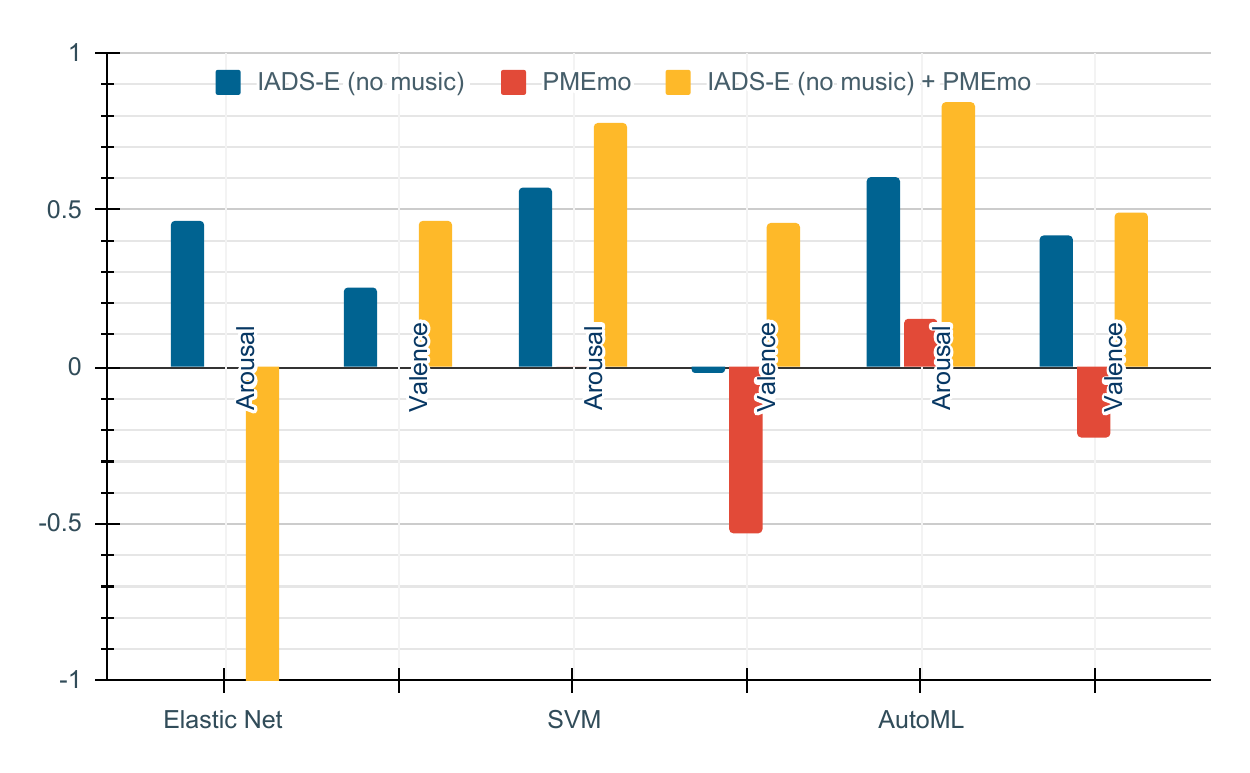}
    \hfill
    \includegraphics[width=0.45\textwidth, trim=0 40 0 0]{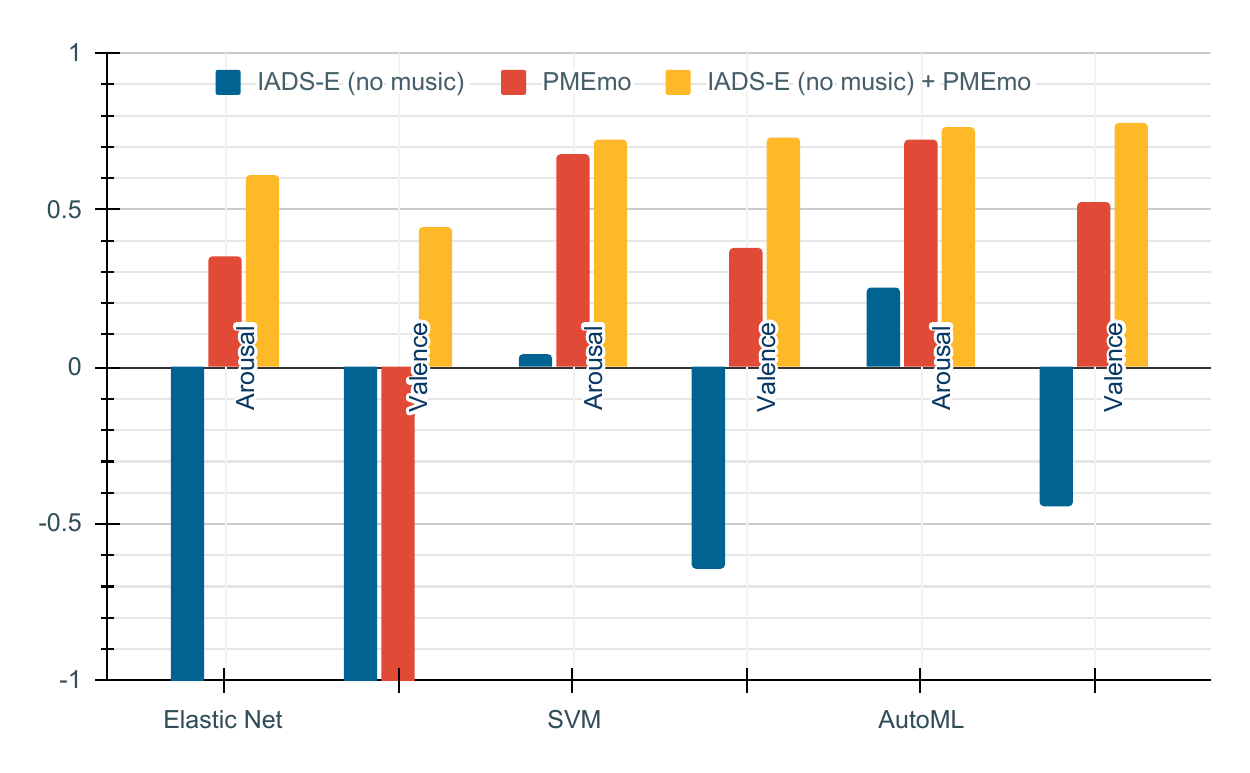}
  \end{center}
  \Description{Image}
  \caption{$R^2$ values according to the various training sets. The test set of the left plot was IADS-E (no music), while for right plot PMEmo. Negative values are truncated at -1.}
  \label{fig:r2}
\end{figure*}

\begin{figure*}[th]
  \centering
  \includegraphics[width=0.85\textwidth]{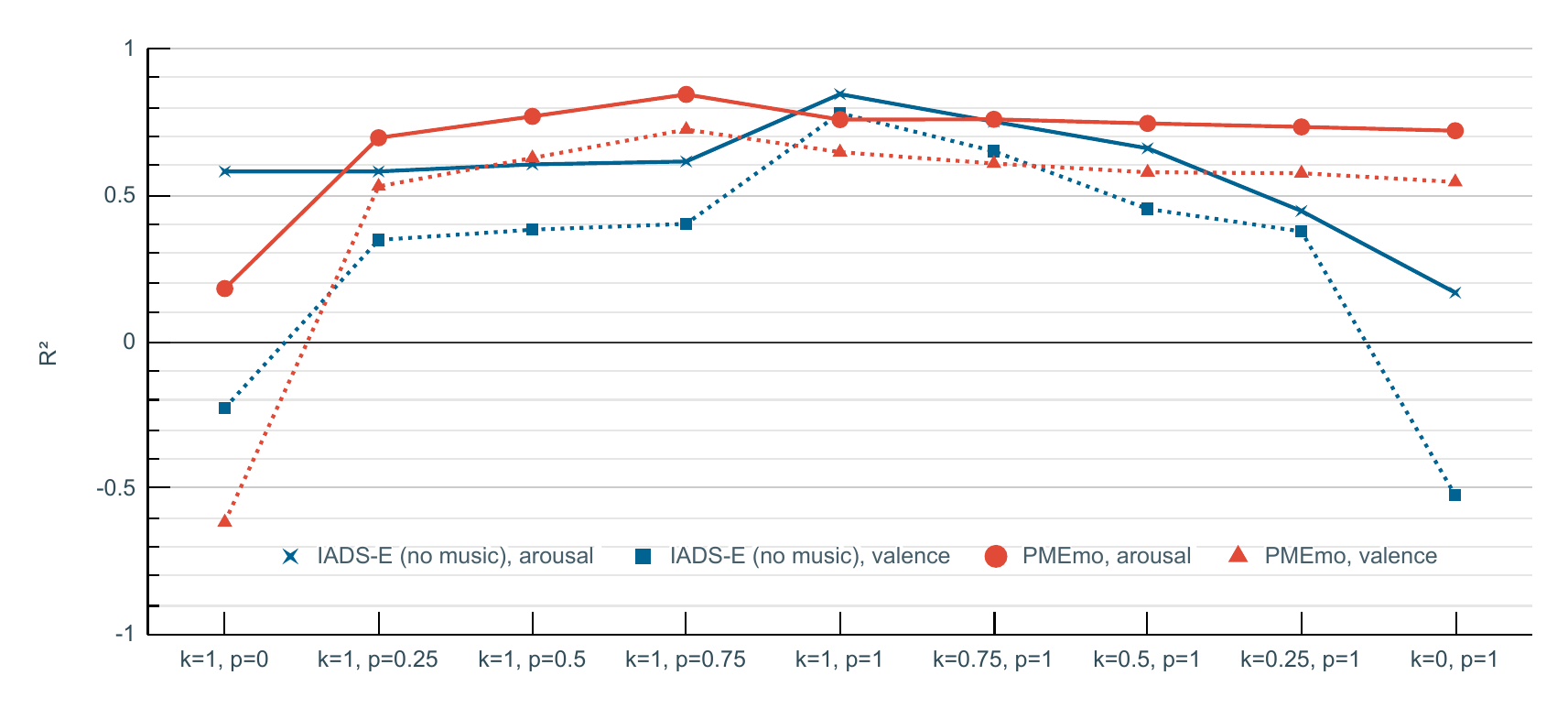}
  \Description{Image}\caption{$R^2$ of the AutoML optimization when different augmentation ratios are used in the train set, i.e. for different values of $k$ and $p$ in the formula $k\times\textit{IADS-E} + p\times\textit{PMEmo}$. Each line represents a different test set, while IADS-E dataset was used without the music samples.
  }
  \label{fig:combinations}
\end{figure*}

We conducted experiments on ElasticNet, SVR, and AutoML, utilizing $p$ and $k$ values from the set $[0, 1]$. This approach only considered augmentation sets generated by the sum of PMEmo and IADS-E. 
The results of the best-performing models for each approach are presented in Table~\ref{tab:rmse}. Additionally, we evaluated the AutoML for various $p$ and $k$ values, as shown in Fig.~\ref{fig:combinations}.

Given that IADS-E includes 170 samples categorized as music, we have excluded them from the experiments, except when comparing with the state-of-the-art, so as to avoid confounding effects when analyzing the inter-relationships between generic sounds and music. This decision was made to ensure the validity and reliability of the obtained results.

\begin{figure*}[th]
  \centering
  \includegraphics[width=0.85\textwidth]{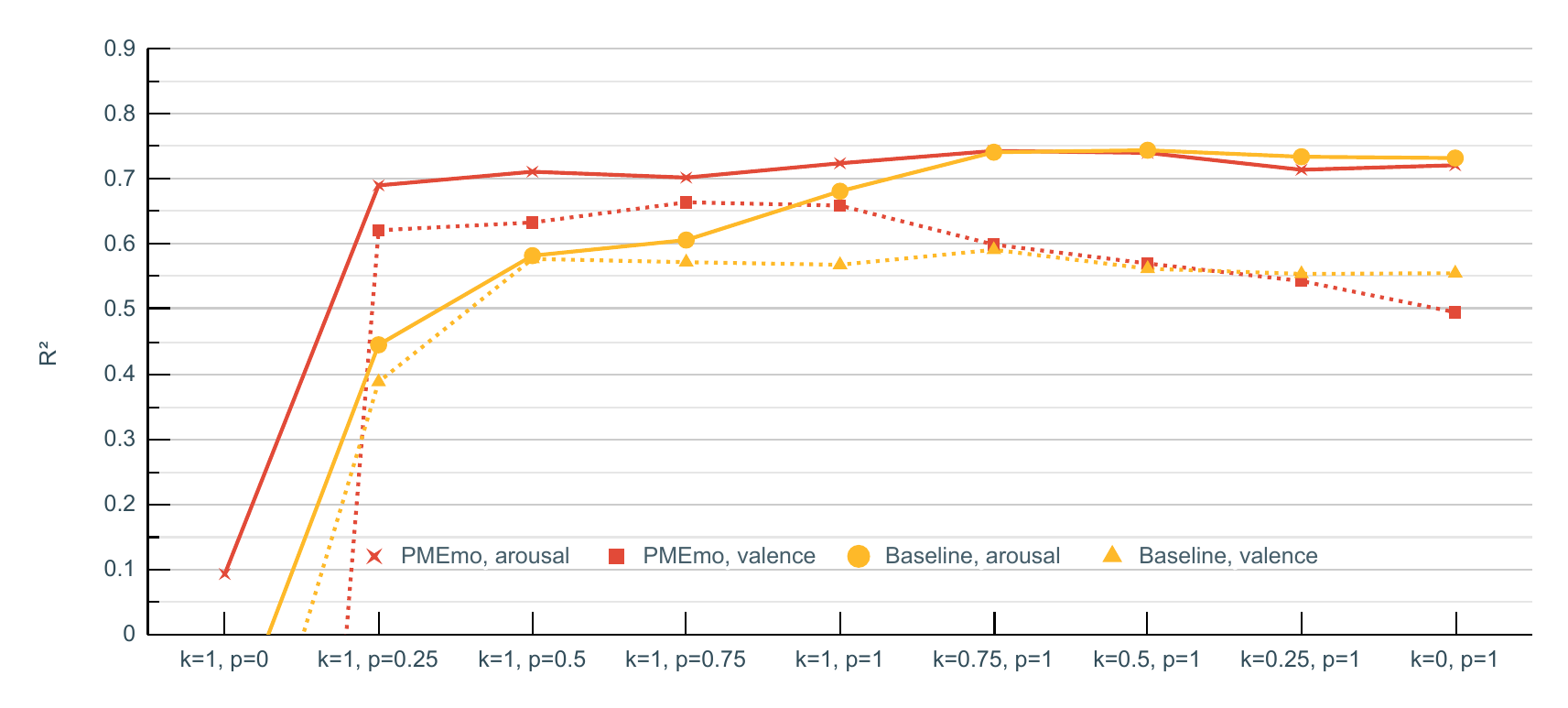}
  \Description{Image}\caption{$R^2$ of the AutoML optimization when different augmentation ratios are used in the train set, i.e. for different values of $k$ and $p$ in the formula $k\times\textit{IADS-E} + p\times\textit{PMEmo}$. Both lines represent $R^2$ scores obtained on the PMEmo validation folds. The baseline is obtained by adding a randomized version of IADS-E to the train set in which the labels were synthesized by uniform random sampling.
  }
  \label{fig:pmemo-with-baseline}
\end{figure*}

We conducted an initial experiment to assess the performance of ElasticNet, SVR, and AutoML on the PMEmo and IADS-E datasets when using either one or both of the datasets. Hence, we executed the AutoML for 8 hours, while the SVM hyper-parameter optimization lasted approximately 4.5 hours and the ElasticNet model 3-5 minutes depending on the size of the train set. The results are presented in Fig.~\ref{fig:r2} and Table~\ref{tab:rmse}. Our findings indicate that all models demonstrate improved performance with the augmentation strategy, particularly AutoML and SVM models. However, the linear ElasticNet model may not fully capture the complexity of the shared space. Therefore, non-linear models might be more appropriate for leveraging this type of augmentation. In fact, ElasticNet does not benefit from the augmentation when predicting emotions conveyed by generalized sounds.

The objective of the second experiment was to assess the performance of AutoML, which was found to be the best-performing approach, when adding a portion of one dataset to another. The experiment ran for 4 hours, during which AutoML searched for the optimal hyper-parameters. The results are presented in Figure \ref{fig:combinations}, where we observe that the most effective augmentation is achieved when both $k$ and $p$ are close to 1. Notably, adding a small amount of music to the training set significantly improves valence prediction for general sounds. Moreover, even without utilizing any sounds from the target dataset, arousal prediction achieves an $R^2$ value greater than 0.15 demonstrating stronger feature sharing across music and general sounds for arousal than for valence.

We also compared the effects of using genuine data versus randomized data in AutoML training. Random labels for the IADS-E dataset were used to train AutoML models for 90 minutes. The performance of these models was contrasted with those trained on the actual IADS-E data, as shown in Figure~\ref{fig:pmemo-with-baseline}. The models trained with random data struggled to find optimal configurations, especially with $k=1$ and $p<1$. In contrast, adding even a small amount of PMEmo data to the real IADS-E dataset improved the models' inference on PMEmo. This observation, underscored in Figure~\ref{fig:pmemo-with-baseline}, illuminates the informational exchange between the IADS-E and PMEmo datasets, suggesting that leveraging commonalities across domains may facilitate the development of superior models.

The third experiment compared the proposed approach with the state-of-the-art. We used AutoML trained on three datasets: PMEmo, IADS-E, and the fully augmented dataset. We also repeated the experiment by including music samples in the IADS-E dataset. For the literature study, we considered all existing works with PMEmo and IADS-E. Standard cross-validation strategies and continuous valence-arousal labels were used. The results are presented in Table~\ref{tab:literature}. We rescaled the RMSE values from \cite{abri2021ComparativeAnalysisModeling} for a fair comparison. Our proposed method surpasses the state-of-the-art due to the model's effectiveness and the augmentation strategy. The ensemble discovered by AutoML is a combination of HistGradientBoostingRegressor models.  However, the resulting solution is slow due to its complexity, resulting in more than 1 hour for the 5-fold cross-validation on a i9-9820X machine with 64GB of RAM. A smaller, more direct model like Support Vector Regressor (SVR) provides improved performance with faster cross-validation time on the same machine (about 30 seconds). Table~\ref{tab:rmse} compares the SVM performance on the PMEmo dataset with the state-of-the-art results in Table~\ref{tab:literature}.

Overall, we argue that the proposed model provides more than satisfactory AER results, where learning from the augmented feature space formed by both music and generalized sounds offers a considerable improvement.



\section{Conclusion}
\label{sec:conclusions}
This paper proposes a new method to improve Audio and Music Emotion Recognition (AER and MER) models. Our approach creates a shared feature space for both types of sounds, allowing for more accurate emotion recognition models in both music and generalized sounds. We extensively validated our proposed strategy and found that non-linear models are necessary for proper modeling of the shared space. Interestingly, our experimental results show that arousal prediction benefits more than valence when learning is done in the shared feature space.


Our strategy is a simple and effective way to enhance the performance of existing models. Lightweight non-linear models like Support Vector Machines can outperform complex neural networks by utilizing the shared feature space.
We believe this method has great potential for various tasks and should be further investigated. In the future, we aim to create a feature space that includes a wide range of data classes to facilitate diverse recognition tasks.

\bibliographystyle{ACM-Reference-Format}
\bibliography{zotero,thesis_references}

\end{document}